# Enhancement of transport critical current density of SmFeAsO$_{1-x}$F$_x$ tapes fabricated by an *ex-situ* powder-in-tube method with a Sn-presintering process

Qianjun Zhang[1], Chao Yao[1], He Lin[1], Xianping Zhang[1], Dongliang Wang[1], Chiheng Dong[1], Pusheng Yuan[1], Shaopu Tang[1], Yanwei Ma[1*], Satoshi Awaji[2], Kazuo Watanabe[2], Yuji Tsuchiya[3] and Tsuyoshi Tamegai[3]

[1]*Key laboratory of applied superconductivity, Institute of Electrical Engineering, Chinese Academy of Sciences, Beijing 100190, China*
[2]*High Field Laboratory for Superconducting Materials, Institute for Materials Research, Tohoku University, Sendai 980-8577, Japan*
[3]*Department of Applied Physics, The University of Tokyo, Hongo, Tokyo 113-8656, Japan*

SmFeAsO$_{1-x}$F$_x$ (Sm1111) tapes were prepared by an *ex-situ* powder-in-tube method with a Sn-presintering process. Scanning electron microscopy revealed apparent difference in microstructure between Sn-presintered tapes and the previously reported polycrystalline Sm1111 bulk, since Sn has reduced FeAs wetting phase and filled the voids between Sm1111 grains. The Sn-presintered tapes showed significant enhanced field dependences of transport $J_c$ compared with Sn-added tapes. A highest transport critical current density ($J_c$) of $3.45 \times 10^4$ A cm$^{-2}$ at 4.2 K and self-field is achieved. Magneto-optical (MO) imaging further confirmed large and well-distributed global and intergranular $J_c$ in Sn-presintered Sm1111 tapes.

*Author to whom any correspondence should be addressed.
E-mail: ywma@mail.iee.ac.cn

Since the discovery of Fe-based superconductors (FBSs),[1] a boom in Fe-based superconductor have been powered up by the mysterious superconducting mechanism and the application potential of very high upper critical fields ($H_{c2}$) above 100 T.[2-5] Among the four main types of FBSs ('1111', '122', '111', '11'), 122 and 1111 are the most promising series for application.[6] Large critical current density ($J_c$) over $10^6$ A cm$^{-2}$ in single crystals and thin films has been discovered in both 1111 and 122.[7-9] 122 has small $H_{c2}$ anisotropy with the ratio γ ~ 2,[10,11] and relatively simple synthetic processes, which make 122 series very promising for application. Meanwhile, 1111 has extremely high $H_{c2}$ values of ~300 T[2] and the highest superconducting transition temperature ($T_c$) of 56 K.[12,13] Recently the $T_c$ of SmFeAsO$_{1-x}$F$_x$ (Sm1111) even reached 58.1 K by using low-temperature sintering.[14] However, a tough challenge for the fascinating 1111 type is obtaining homogeneous superconducting phase without impurities, which is the thorniest issue for 1111 type application.

The first FBS wires were soon fabricated from Sm1111[15,16] and Sr$_{0.6}$K$_{0.4}$Fe$_2$As$_2$[17] by powder-in-tube (PIT) method. In the past three years, FBS wires and tapes have achieved tremendous progress.[18-23] The transport $J_c$ of 122 tapes has reached >$10^4$ A cm$^{-2}$ at 4.2 K and 10 T in the latest report. However, at present the $J_c$ of 1111 tapes is relatively low because of the difficulty in controlling fluorine loss and detrimental impurities during sintering.[24] Wang *et al*[23] reported a transport $J_c$ value of 2.2 × $10^4$ A cm$^{-2}$ (at 4.2 K and self-field) by Sn-added Sm1111 tapes, which is the highest $J_c$ among the 1111 wires and tapes so far. Sn addition can effectively enhance the $J_c$ value in weak field, but as the field increases the $J_c$ value drops quickly. Therefore, the fascinating 1111 having the highest $T_c$ and $H_{c2}$ among FBSs urgently needs an improvement on wires and tapes.

In this work, we demonstrate a fabrication process of Sm1111 tapes. Sn addition is still adopted but together with a Sn-presintering process before wire fabrication. The Sn-presintering process resulted in an significant enhancement of global supercurrent. A transport $J_c$ of 3.45 × $10^4$ A cm$^{-2}$ (at 4.2 K and self-field) is achieved which is the highest value ever reported among 1111 wires and tapes.

Mixtures of Sm, Fe, As, Fe$_2$O$_3$, SmF$_3$ powders in molar ratio of

0.92:0.44:1:0.28:0.08 according to the nominal composition SmFeAsO$_{0.84}$F$_{0.24}$ were well ground using a ball milling in Ar atmosphere for 10 h. The excess fluorine is added to compensate the fluorine loss during synthesis. Then the mixed powders were sealed in an iron tube by arc welding and then heat treated at 1100 ℃ for 30 h to get Sm1111 precursors. The Sm1111 precursors were ground and Sn powders were added and well mixed. The Sn-added precursors were sealed in an iron tube by arc welding again. The tube was then put into a 1100 ℃ furnace for 2 minutes followed by a furnace cooling immediately. The Sn-presintered precursors were ground to powders again and sealed in iron tubes. Then the filled tubes were swaged and drawn down to wires. The wires were cold-rolled to thin tapes with a thickness ~0.6 mm. Finally, the tapes were heat treated at 1100 ℃ for 40 seconds.

The field dependence of transport $I_c$ at 4.2 K for tapes was measured at the High Field Laboratory for Superconducting Materials (HFLSM), by a standard four-probe resistive method, with a criterion of 1 μV cm$^{-1}$. Magnetization measurements were performed using a Quantum Design physical property measurement system (PPMS). Phase composition was characterized by x-ray diffraction (XRD; Rigaku RINT2000) analysis with Cu-K radiation. The microstructures were measured by scanning electron microscopy (SEM). Magneto-optical (MO) images were obtained at The University of Tokyo.

Fig. 1(a) displays the XRD pattern of the pure Sm1111 precursor powder. Sm1111 is confirmed as the primary crystalline phase. The main impurities include SmAs and SmOF. After the Sn-presintering process, Sn and FeSn$_2$ phases are identified in the powder, as presented in Fig. 1(b). The *a* lattice parameter is 0.3927 nm, indicating the precursor powder still has a proper fluorine concentration after Sn-presintering process. The XRD patterns of Sn-presintered tapes and Sn-added tapes (Fig. 1(c) and (d)) were measured after peeling away the sheaths on one side, and hence their diffraction peaks are not as sharp as precursor powders. In the tapes processed by both methods, the main impurities are Sn, FeSn$_2$, SmAs and SmOF, similar to the Sn-presintered powder.

The temperature dependences of normalized resistivity for pure Sm1111

precursor, 27 wt.% Sn-presintered precursor and final tape by 27 wt.% Sn-presintered precursor are presented in Fig. 2. The pure Sm1111 precursor synthesized in the first step has a $T_c^{onset}$ of 54.2 K, which is typical for the fluorine doped Sm1111 polycrystalline bulk. The $T_c^{onset}$ for 27 wt.% Sn-presintered precursor is 53.3 K. The Sn-presintering process caused a slight decline on $T_c$ but remained the sharp transition curve similar to the pure precursor. The tape fabrication processes caused the most significant $T_c$ degradation. A tail shaped transition is observed in tape samples with $T_c^{onset}$ ~ 51.1 K and $T_c^{zero}$ ~ 41.7 K, which might be caused by the damage to composition and structure of Sm1111 crystals during tape fabrication.

Fig. 3 shows the field dependence of transport $J_c$ at 4.2K for Sm1111 tapes by different processes. In general, Sn-presintered samples show a drastic enhancement of $J_c$ in full range of magnetic fields compared with Sn-added samples. The 27 wt.% Sn-presintered Sm1111 tapes achieved the highest $J_c$ value of $3.45 \times 10^4$ A cm$^{-2}$ (at 4.2 K and self-field), which is 57 % higher than the highest $J_c$ record[23] for 1111 wires and tapes, with $I_c$ of 234 A and a transverse cross-section area of 0.68 mm$^2$. An even more interesting result is that in fields > 1 T, 10 wt.% and 27 wt.% Sn-presintered samples have much milder degradation than Sn-added samples. Achieving a high $J_c$ in strong field is an essential demand since FBSs are aiming at high-field applications.

Fig. 4(a) shows an optical image of the polished transverse cross-section of the 27 wt.% Sn-presintered tape. The core and sheath remained in fine shape after drawing and rolling, but a crack can be observed in the middle of the core. Fig. 4(b) is an optical image of cross-section near the boundary between Fe sheath and superconducting core. Fig. 4(c) is the back-scattered electron (BSE) SEM image of polished surface of the core parallel to the tape plane. The BSE-SEM images of transverse cross-section of the 27 wt.% Sn-presintered tape are presented in Fig. 4(d)-(f). No reacting layer between Fe sheath and superconducting core was observed in Fig. 4(b) and (d). Energy dispersive x-ray spectroscopy (EDX) area mapping results for Fig. 4(e) are displayed in Fig. 5. The dark area, white area and grey area in BSE-SEM images have been identified by EDX analysis as FeAs, SmAs and mixture of Sm1111 and Sn, respectively. From careful observation of Fig. 4(e) and (f), a brighter grey contrast and a darker grey contrast can be distinguished from the grey area. As EDX mapping result (Fig. 5) shows the area with more Sn distribution (Area A) looks brighter in BSE-SEM image than the area with less Sn distribution (Area B), we can speculate that the brighter grey contrast in Fig. 4(f) corresponds to Sn or $FeSn_2$ and the darker grey contrast corresponds to Sm1111. Therefore, in Sn-presintered tapes Sn and $FeSn_2$ were mixed with Sm1111 grains and filled the voids between Sm1111 grains. This is very different with the SEM image of the pure polycrystalline Sm1111 bulk.[24-26] In polycrystalline Sm1111 bulk, amorphous FeAs wetting phase exists between Sm1111 grains as the most detrimental impurity blocking intergranular supercurrent. Previous reports also suggest In and Sn additions can reduce FeAs impurity.[24,27] The reduction of FeAs and better grain connectivity by Sn-presintering process could be possible reasons for the $J_c$ enhancement.

Fig. 6 presents the field dependences of magnetic $J_c$ at 5 K for 27 wt.% Sn-presintered tapes and 27 wt.% Sn-added tapes calculated from the hysteresis loops shown in the inset using Bean model. The magnetic $J_c$ values of Sn-presintered tapes show enhancement in full range of magnetic fields, which is consistent with the result from transport measurement. For Sn-added Sm1111 tapes, the transport $J_c$ drops more quickly than the magnetic $J_c$ as the field increases, but for Sn-presintered tapes the difference between transport $J_c$ and magnetic $J_c$ becomes less. This could imply the Sn-presintered tapes allow more global supercurrent than Sn-added tapes because the difference between transport $J_c$ and magnetic $J_c$ comes from the local and intragranular supercurrent contribution.

Enhancements of intergranular $J_c$ are directly visualized by magneto-optical (MO) imaging. Fig. 7(h) shows the optical image of a sample from the core of the 27 wt.% Sn-presintered tapes. The sample has a thickness of 0.1 mm. The MO images taken at 5 K with the field increasing to 200 Oe (Fig. 7(a)) and 400 Oe (Fig. 7(b)) indicate the fields flux started penetration into the sample from the edges, but the flux-free region still existed in the center of the sample until 400 Oe. A nearly uniform rooftop pattern of magnetic flux density was produced by global current flow throughout the entire sample. Fig. 7(c)-(g) depict MO images of the sample in the remnant state at different temperatures after applying different fields that were subsequently reduced down to zero. The bright regions correspond to the trapped flux in the sample. These images show great enhancement compared to the previously reported MO images of 1111 polycrystalline bulks and tapes.[23,28,29] In the past studies, the bright intensities are restricted in many small regions, implying that the intergranular $J_c$ is much smaller compared with the intragranular $J_c$. In Fig. 7(c)-(g), the trapped flux shows a global distribution in all conditions. This improvement indicates the sample by the Sn-presintering process has a much larger intergranular $J_c$ than before, which corroborates the result of transport $J_c$ and magnetic $J_c$ measurement.

Transport $J_c$, magnetic $J_c$ and MO imaging measurement have identified the significant enhancement of global $J_c$ in Sm1111 tapes by Sn-presintering process. However, the $J_c$ values in strong fields are still very low when compared with 122 tapes by PIT method. A major drawback in Sm1111 tapes is that there exist too many impurities in the superconducting core as shown in SEM images. To find a route to obtain pure Sm1111 phase with appropriate fluorine concentration is the most critical challenge. The recent studies on low-temperature sintering to synthesize Sm1111[14,30] and Ba122[31] bulks provided an idea to prevent FeAs liquid phases from forming as the reaction temperature was below the liquidus line in FeAs phase diagram. However, many unreacted reactants remain in Sm1111 bulks after a long sintering, as the reaction rate at low temperature is relatively low. Another drawback in the present Sm1111 tapes is that the tapes cannot endure long-time sintering after drawing and rolling, as the severe fluorine loss will occur when post-rolling sintering lasts longer than 80 seconds.[24] In 122 tapes, a high transport $J_c$ was usually obtained after a post-rolling sintering longer than 30 minutes.[32] Therefore, to develop a new technique to make the Sm1111 tapes able to endure longer post-rolling sintering could lead to a breakthrough in Sm1111 tapes.

In summary, a series of Sm1111 tapes were fabricated by an *ex-situ* PIT method with a Sn-presintering process. The 27 wt.% Sn-presintered Sm1111 tapes achieved a transport $J_c$ of $3.45 \times 10^4$ A cm$^{-2}$ at 4.2 K and self-field. SEM studies revealed that Sn has displaced FeAs and existed between Sm1111 grains. It is very different from the observations of polycrystalline Sm1111. Magnetic $J_c$ measurement and MO images indicate the global and intergranular $J_c$ of Sm1111 tapes have been improved significantly by the Sn-presintering process.

This work was supported by the National "973" Program (Grant No. 2011CBA00105), National Natural Science Foundation of China (Grant Nos. 51025726, 51320105015 and 51202243), and China-Japan Bilateral Joint Research Project by NSFC and JSPS.

# Captions

Fig. 1. XRD $\theta$-$2\theta$ pattern for (a) pure Sm1111 precursor powder; (b) Sn-presintered Sm1111 powder; (c) Sn-presintered Sm1111 tape; (d) Sn-added Sm1111 tape. The pure Sm1111 precursor shows good quality. The Sn-preintered powder, Sn-added tapes and Sn-presintered tapes include similar impurities.

Fig. 2. Temperature dependences of normalized resistivity for pure Sm1111 precursor, 27 wt.% Sn-presintered precursor and final Sm1111 tape by 27 wt.% Sn-presintered precursor. Inset shows the enlarged view near the superconducting transition temperature.

Fig. 3. Magnetic field dependence of transport $J_c$ at 4.2 K for Sn-added Sm1111 tapes and Sn-presintered Sm1111 tapes with 10 wt.% and 27 wt.% Sn quantities. The 27 wt.% Sn-presintered tapes achieved the highest $J_c$ value of $3.45 \times 10^4$ A cm$^{-2}$ at 4.2 K and self-field. The Sn-presintered tapes show $J_c$ enhancement compared with the Sn-added tapes.

Fig. 4. (a) and (b) optical image of transverse cross-section of 27 wt.% Sn-presintered tape. BSE-SEM image of the polished (c) cross-section parallel to the tape plane, (d)-(f) transverse cross-section of the tape. SmAs and FeAs impurities was identified by EDX. The grey area was identified as a mixture of Sm1111 and Sn.

Fig. 5. EDX area mapping for Fig. 4(e). The dark contrast, white contrast and grey contrast is identified as FeAs, SmAs and mixture of Sm1111 and Sn. Area A contains more Sn than Area B. Accordingly, in BSE-SEM, Area A has a brighter contrast than Area B.

Fig. 6. Field dependence of magnetic $J_c$ at 5 K for (a) 27 wt.% Sn-presintered Sm1111 tapes; (b) 27 wt.% Sn-added Sm1111 tapes. Sn-presintered tapes show larger magnetic $J_c$ values in all range of fields compared with the Sn-added tapes. The inset shows the hysteresis loops of the samples.

Fig. 7. Magneto-optical images of superconducting core sample from 27 wt.% Sn-presintered Sm1111 tape at (a) 200 Oe, (b) 400 Oe after zero-field cooling

down to 5 K. Differential MO images in the remanent state after cycling the field up to (c) 200 Oe, (d) 400 Oe, (e) 600 Oe at 5 K; (f) 200 Oe at 15 K; (g) 25 Oe at 30 K. The images show large global $J_c$ in the sample.

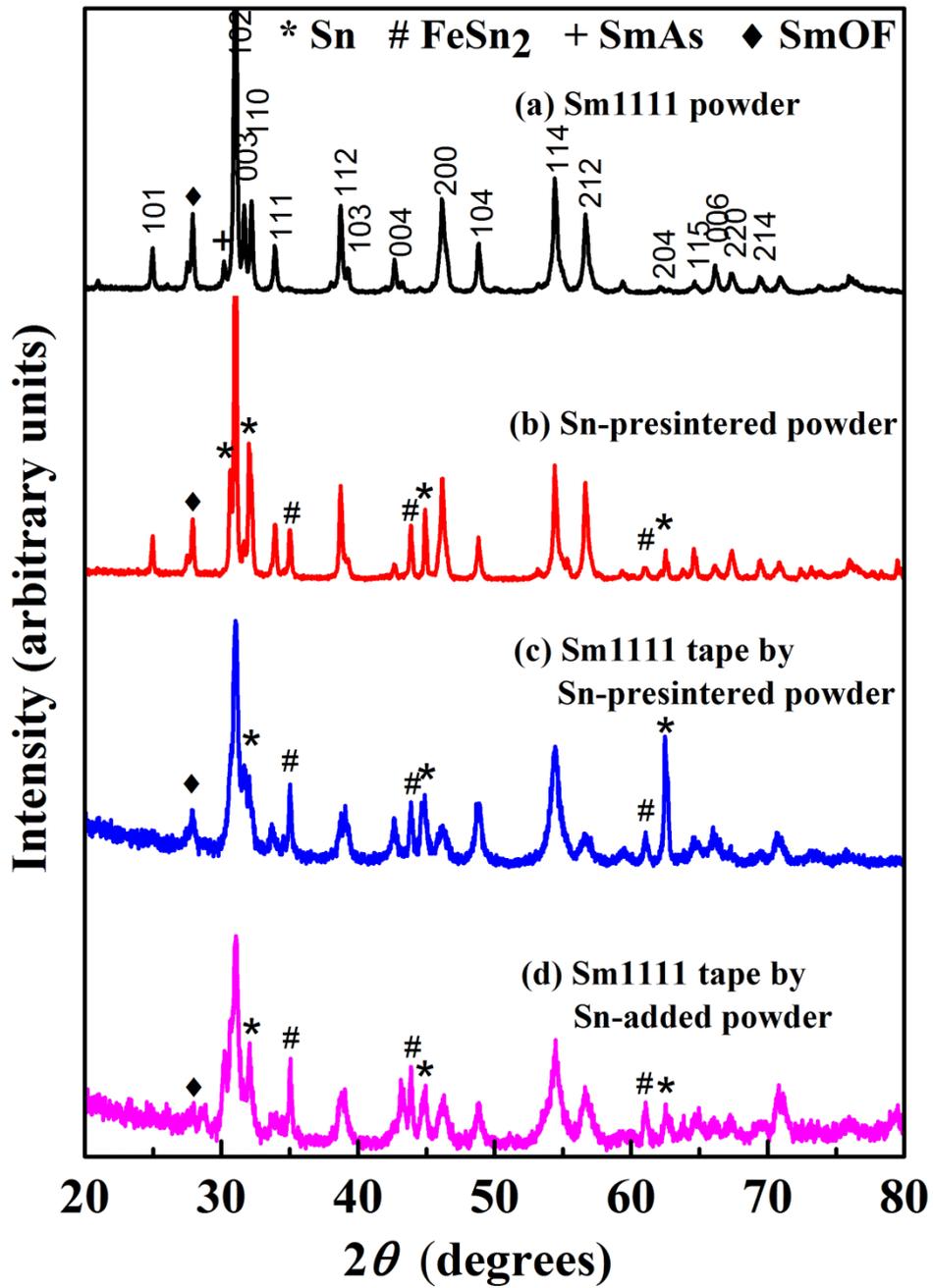

Fig. 1. Q.J. Zhang *et al*.

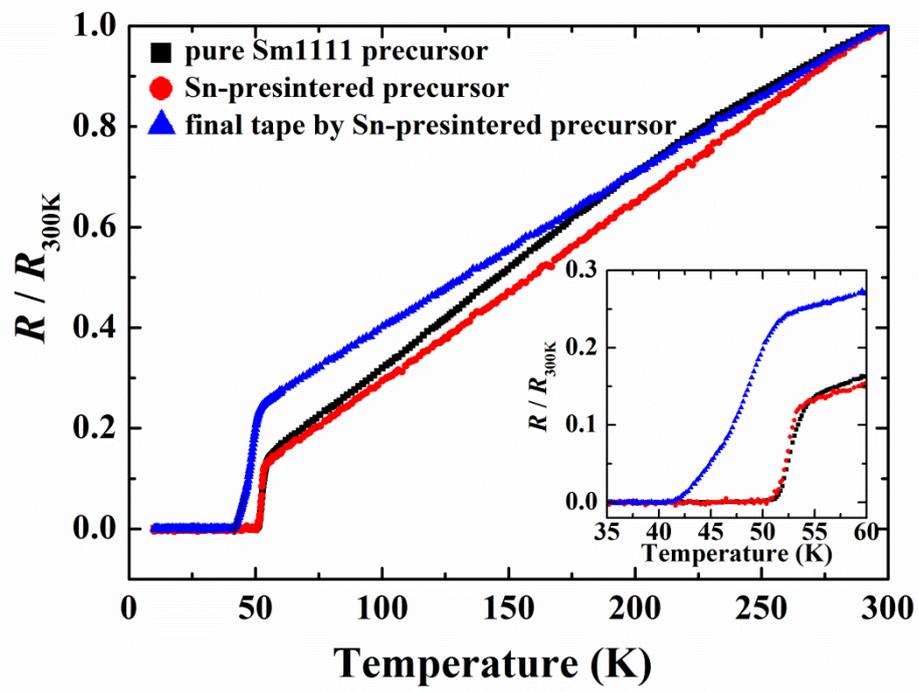

Fig. 2. Q.J. Zhang *et al*.

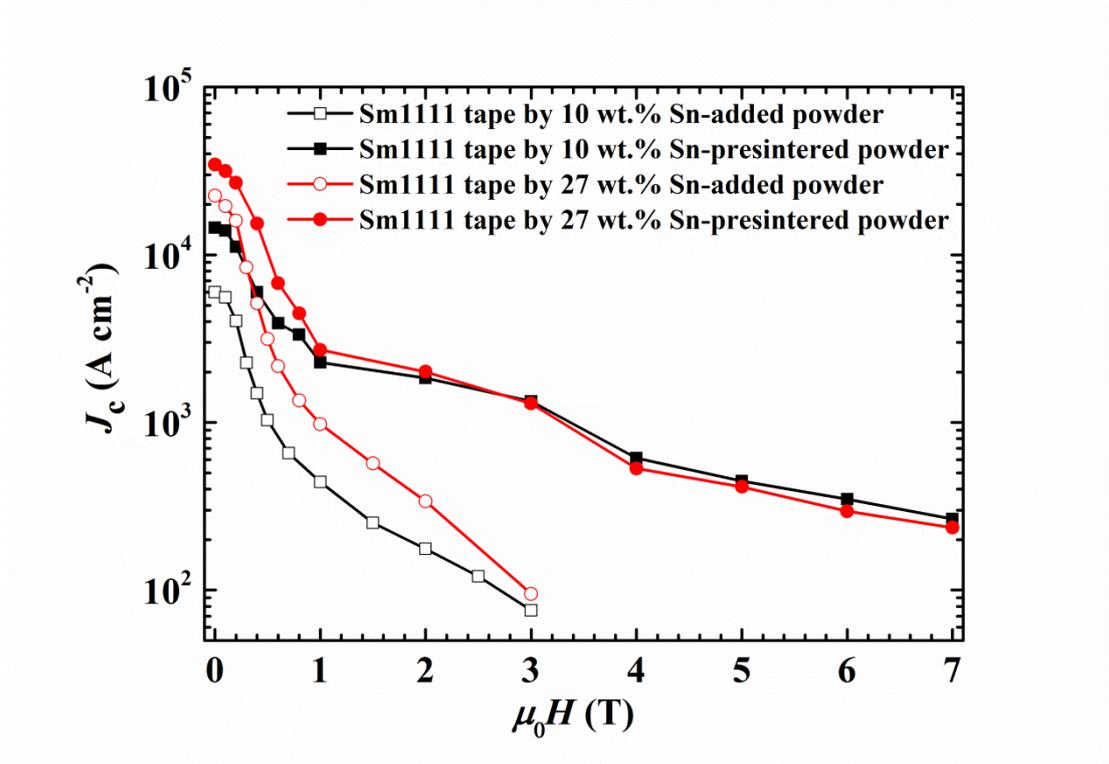

Fig. 3. Q.J. Zhang *et al*.

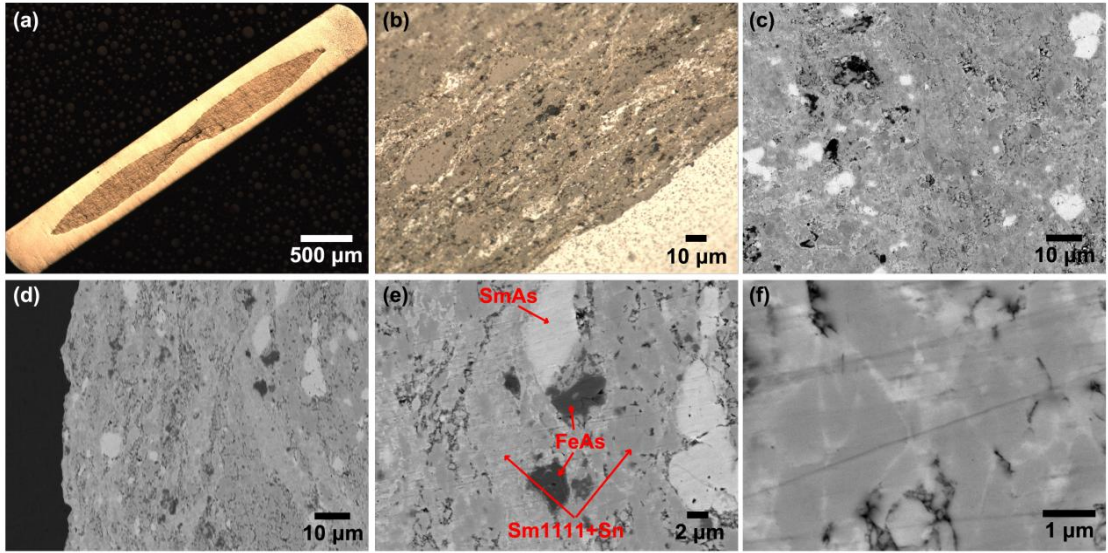

Fig. 4. Q.J. Zhang *et al*.

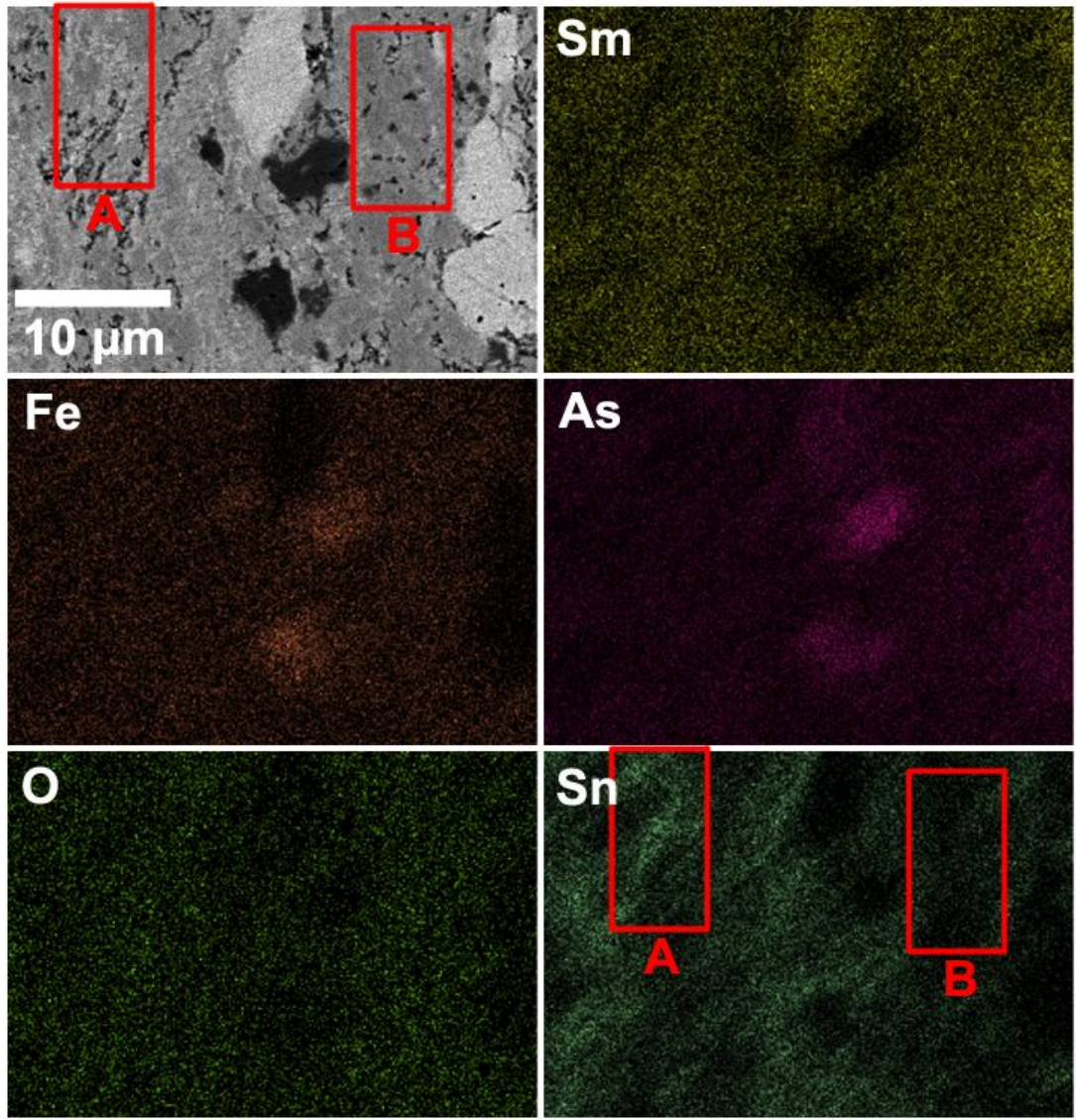

Fig. 5. Q.J. Zhang *et al*.

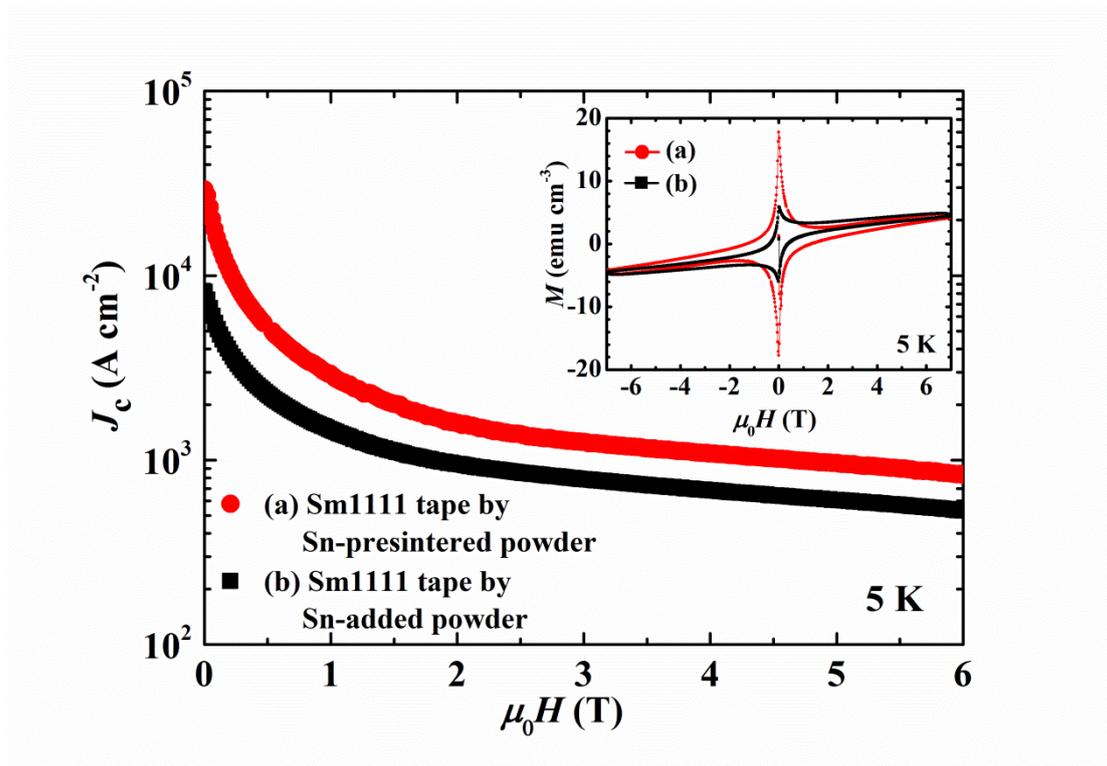

Fig. 6. Q.J. Zhang *et al*.

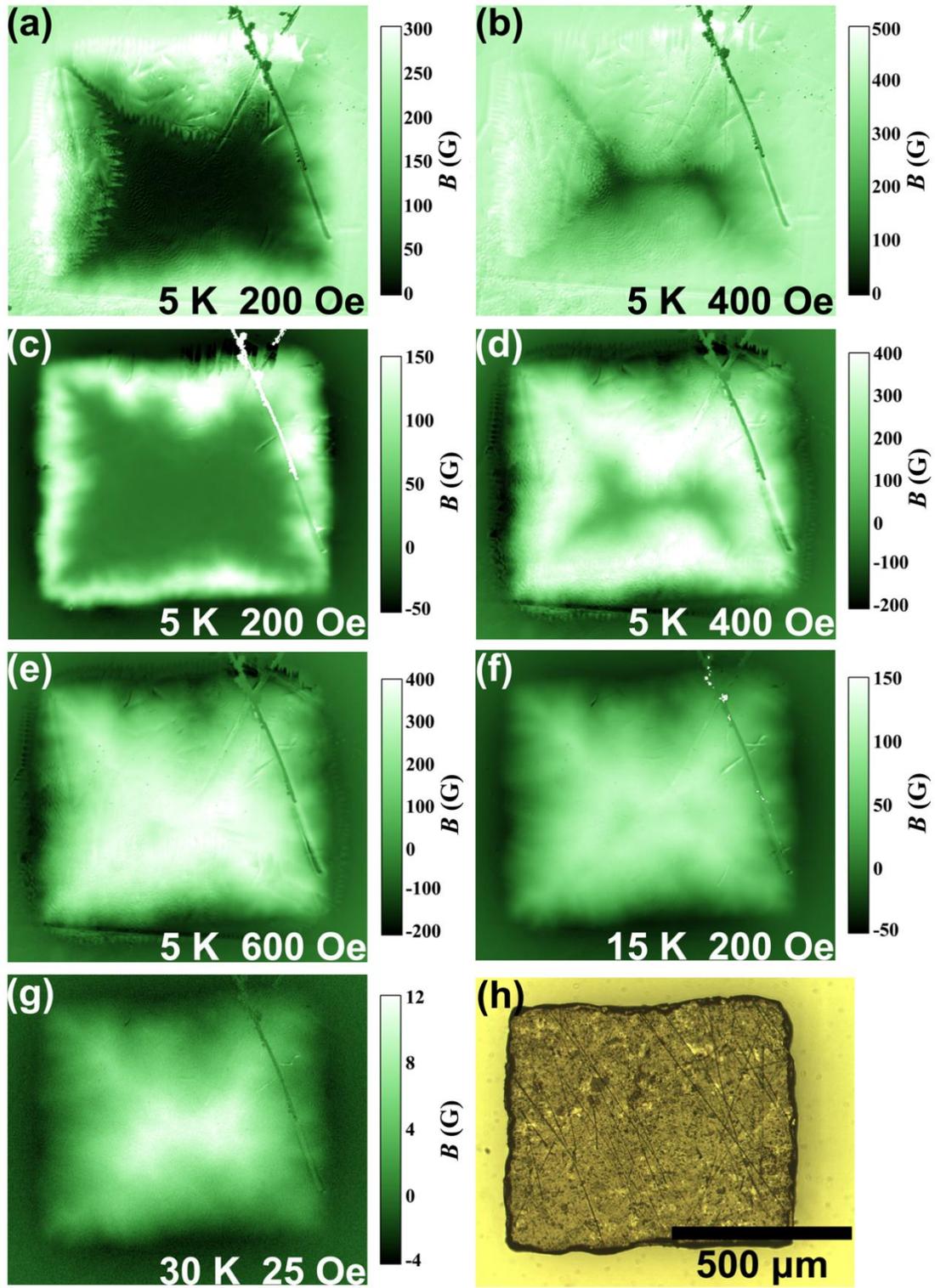

Fig. 7. Q.J. Zhang *et al*.